**Fractal-Mound Growth of Pentacene Thin Films: A Novel Morphology**

Serkan Zorba

Department of Physics and Astronomy, Whittier College, Whittier, California, 90608

Yonathan Shapir, and Yongli Gao

Department of Physics and Astronomy, University of Rochester, Rochester, New York 14627

**Abstract**

The growth mechanism of pentacene film formation on $SiO_2$ substrate was investigated with a combination of atomic force microscopy measurements and numerical modeling. In addition to the diffusion-limited aggregation (DLA) that has already been shown to govern the growth of the ordered pentacene thin films, it is shown here for the first time that the Schwoebel barrier effect steps in and disrupts the desired epitaxial growth for the subsequent layers, leading to mound growth. The terraces of the growing mounds have a fractal dimension of 1.6, indicating a lateral DLA shape. This novel growth morphology thus combines horizontal DLA-like growth with vertical mound growth.

In recent years, organic semiconductor materials have demonstrated their potential as a new class of electronic materials that offers numerous new and important applications [1, 2]. One of the most important parameters of these materials is the mobility with which the charge carriers move in the device, which in turn determines the speed of the device. It has been shown that the mobility depends crucially on the morphology of the organic semiconductor that is being used as the active material in the device [3, 4]. However, only recently have there been reports on the growth mechanisms of organic semiconductors [5, 6].

Pentacene has been studied as a model organic semiconductor in the field of organic electronics because of its high mobility that is comparable to that of amorphous silicon [7, 8]. The reason for pentacene's high mobility is its forming highly ordered molecular crystalline thin films as confirmed with X-ray diffraction studies [3]. It has also been shown that having few large single crystals of organic semiconductor instead of many small ones improves mobility due to the reduced number of grain boundaries that normally hamper the charge transport [3, 4, 9]. To this end, Jackson and his co-workers treated the $SiO_2$ substrate with octadecyltriclorosilane (OTS) before depositing pentacene on it. This OTS pre-treatment produced larger pentacene single crystals and, as a result, the mobility was improved [10]. Heringdorf and co-workers succeeded in growing large single crystals of pentacene on Si (001) surface in the order of 0.1 mm [5]. Pentacene has also been shown to form ordered films on $SiO_2$ and other substrates [11]. Common to the orderly growth of the pentacene thin films is the formation of DLA-type fractal branches with pyramidal upward growth. This growth is an intriguing one. New nucleation centers form on the already existing DLA-like monolayer islands before the latter get a chance to coalesce and form a complete epitaxial layer. This in turn limits the size of the single crystal. The mechanisms behind such peculiar growth of pentacene have not been fully studied. So far, some efforts have been made only to reveal the fractal character of the growth. Its sub-monolayer evolution was characterized by a growth mechanism of the DLA type, modified to take into account of overlapping diffusion fields as more molecules land on the substrate [5].

The recent application of scaling theories to understand the mechanisms involved in the evolution of organic/inorganic interfaces is noteworthy, considering the fact that the major efforts in applying such theories have been so far focused on the study of the growth of inorganic thin films [6, 12, 13]. More recently, a study on the initial monolayer growth of pentacene on reduced and oxidized silicon surfaces was reported [14]. However, no study was made in order to fully describe the growth of the subsequent layers of orderly pentacene growth on oxidized silicon substrates as used in organic thin film transistors (OTFT). Especially, the disruption of the epitaxial growth of pentacene after the first monolayer and the resulting upward growth has been a major concern but not yet fully understood. Such a novel morphology, or irregularly-shaped mounds, has never been observed experimentally in any other organic system before, nor was it theoretically suggested.

In this letter, we present our explanation for this novel morphology. We found that it arises from the conjunction of two basic growth processes: DLA-like growth in the horizontal direction, and mound growth in the vertical direction. Our conclusion is based on detailed atomic force microscopy (AFM) measurements and modeling of the process with numerical simulations. To our knowledge, this is the only example in which the DLA and mound growth coexist in the growth of an organic material. Understanding such novel morphologies is critical for the kinetic roughening and thin film studies in general, especially in systems with relatively weak interactions such as organic molecular semiconductors. Furthermore, the implications of the observed mechanisms may provide guidelines for future improvement of device characteristics through morphological modifications, as well as understanding the limitation of the conductance in organic thin-film transistors to the first few monolayers from the organic-gate-oxide interface [15].

As-received pentacene, purchased from Aldrich Chem. Co., was thermally evaporated on $SiO_2$ substrate held at room temperature at a base pressure of about $3 \times 10^{-6}$ torr with about 0.5 Å/s deposition rate. Different thicknesses were obtained by masking successive parts of the sample at the corresponding time values. Using this gradual thickness method, we obtained all the different thicknesses in one experiment which left us with a thin film pentacene staircase, each step corresponding to a different thickness and, hence, different time. This allowed us to examine both the spatial and temporal correlations of the surface roughness using atomic force microscopy (AFM) with precisely identical conditions. The surface morphology measurements were done using a Topometrix Explorer AFM in the tapping mode in air. A planar background subtraction was made for all the data to correct for the tilt of the sample with respect to the scanning plane. Computer simulations were done with True Basic.

Shown in the top panel of Fig. 1 are some representative tapping mode AFM images of pentacene thin films evaporated on $SiO_2$ substrate. At low coverages we observed a few round islands which gradually took a DLA-like shape as the coverage increased. However, new monolayers start nucleating without waiting the underlying monolayer(s) to form completely. This could be seen in the top panel of Fig. 1. This DLA-like growth has been recently investigated in the literature, and our results agree with them. However, a closer look at the AFM images of the pentacene film in Fig. 1 reveals that the ordered growth of pentacene is a very peculiar one that incorporates a number of different growth mechanisms, not just fractal growth as caused by DLA mechanism. It can be seen from the top panel of Fig. 1 that pentacene forms a mixture of both regular and irregular-sized mounds on $SiO_2$ surface. In addition, one can readily observe the monolayer structures forming pyramid-like mounds. In fact, the average step size of the monolayers was found to be 16 ± 1 Å, which matches closely with the length of a pentacene molecule (see Fig. 2) [3]. Moreover, these terraced mounds are in a shape that bespeaks fat-DLA growth. DLA forms via diffusion of molecules through pure random walk and their sticking to a cluster to create irregularly branched and chaotic patterns. Fat DLA forms when the particles are allowed to diffuse on the boundary of the cluster to increase the number of nearest neighbors [16]. Indeed, our fractal dimension analysis revealed a dimension of about 1.6, which agrees very closely with that of a DLA surface [17]. In addition, the Fourier transform of the surface revealed a ring-like behavior, a sign of mound growth (see Fig. 3) [18]. As a result, what one has as the dominant growth mechanisms behind pentacene's growth seems to be an interplay between the DLA-type growth and mound growth.

There have been a number of epitaxial growth experiments recently, which have shown pyramid-like mound morphology in inorganic systems [19, 20]. This phenomenon is mainly attributed to a growth instability caused by the step (or Schwoebel) barrier that creates a bias in the upward and downward diffusions in favor of the former. In the presence of a step-edge barrier on a terrace, the newly landing molecules are prevented from hopping down the edge, which in turn spawns an uphill current [17, 21]. This will result in the formation of regular pyramid-like mounds. The power spectrum of such a surface will be ring-like, indicating a length selection in real space.

To further confirm whether it is really two mechanisms that dominate the growth of pentacene, we have performed Monte Carlo simulations. We used two exclusive models. (i) The standard DLA model with surface-curvature-dependent sticking probability scheme as was first shown by Vicsek [16]. However, due to the weak van der Waals interactions among pentacene molecules, we have allowed a small probability for the particles to stick to any site (thus also acting as seeds) while they diffused randomly on the surface. (ii) The Schwoebel barrier model. The simulations were carried out on a square lattice. This is a convenient approximation to mimic the herringbone structure [14] of ordered pentacene as was also employed by Heringdorf et al [5] in their simulation of the first monolayer growth of the latter. In such a structure, a molecule would have four nearest neighbors and four next-nearest neighbors. The local curvature of the surface was characterized by the number of neighboring particles. A range of neighbor criterion was studied, from having just one or two nearest-neighbors to counting next-nearest neighbors. Low coordination numbers produce very irregular and thin DLA branches inconsistent with the AFM topography. Likewise, higher sticking coefficients produce too many nucleation centers reducing the diffusion length of the particles to form any DLA behavior. To obtain the Monte Carlo results shown in the lower panel of Fig. 1, we used ≥ 3 nearest neighbors and sticking coefficient of $10^{-5}$.

To account for the Schwoebel effect we prohibit a particle from moving down the step of a terrace, thus inducing an upward diffusion bias which leads to mound growth. Bottom panel of Fig. 1 shows some of the typical results we have obtained. To the eye, the real morphology data and the simulation morphology data resemble each other closely. In fact, one can easily distinguish the DLA-type formation in the initial coverages (lower panel Fig. 1(c)). With more particles landing, due to the uphill diffusion bias driven by the Schwoebel barrier effect, newer DLA-type monolayers formed on top of existing ones in a much faster fashion. As a result, as can be seen in the simulation figures, a novel morphology was generated that closely mimics that of pentacene. In both sets of data, individual monolayer terraces can easily be identified. DLA with almost dendritic patterns can also be seen. One has to note here that the pentacene branches observed are not exactly like the usual long and chaotic DLA branches. However, the fractal dimension we measured (by comparing the area to their perimeter) was exactly that of DLA model. Furthermore, DLA clusters don't have to be always dendritic with long and disordered branches. DLA fingers can be wide, compact, and even nearly regular depending on the surface-curvature criterion [16] and diffusion length [17].

Further quantitative understanding of the growth mode can be obtained by analyzing the scaling exponents of the system. These exponents provide key information about the growth mechanism, characteristics, and universality class [17, 22]. The scaling approach [17, 23] was introduced primarily to study self-affine surfaces. In such surfaces, height-height correlation function, $C(r, t)$, is used to characterize the growth. The height-height correlation function is the rms fluctuations in the height difference between two points separated by a distance r at time t. It behaves as $C(r, t) \sim \rho r^{\alpha}$ for distances less than the correlation length $\xi(t)$, where $\alpha$ is the roughness exponent. For $r > \xi(t)$, it saturates to the value of the roughness which increases with time as $t^{\beta}$, where $\beta$ is the growth exponent. These exponents are used to describe mound growth as well. In that case $\beta$, sometimes replaced by n when related to mounds, describes the mound height growth and is larger than 0.25 [17, 19]. $\xi(t)$ is then the lateral size of the mounds. The roughness exponent is trivially $\alpha = 1$ if a single slope is selected, which is almost always the case for regular (circular) mounds. If the mounds are not circular, the slope is not uniform and $\alpha$ is then an "effective" roughness exponent smaller than one (since the smaller slopes become more important at large distances). We will use this "effective" roughness exponent $\alpha_{eff}$ below only as another available tool to characterize the non-circular pentacene mounds.

The height-height correlation function C(r) of both the real and simulation data was calculated for different thicknesses. We plot, in the left panel of Fig. 4, C(r) as a function of distance r, and the saturated $C_{sat}$ (r) as a function of time in the right panel of Fig. 4 for both the real (top panel) and simulation (bottom panel) data. For small distances, C(r) changes linearly with r, and saturates beyond the characteristic length $\xi(t)$. The characteristic length is about the size of individual islands. The fact that the former is not larger than the latter suggests that the surface is not self-affine in the usual sense. It has a different self-affinity: mound growth rescaling in the vertical direction and DLA-type rescaling in the lateral direction [24].

The calculated $\alpha_{eff}$ and $\beta$ values for the real data are $0.57 \pm 0.03$ and $0.27 \pm 0.03$, respectively. The simulation values for $\alpha_{eff}$ and $\beta$ are $0.48 \pm 0.07$ and $0.29 \pm 0.01$, respectively. The close agreement of these two sets of numbers with each other further corroborates quantitatively our earlier assessment that the morphology of the pentacene on $SiO_2$ is dominated by two different growth mechanisms, diffusion and step-edge barrier effect. Their interplay results in a novel and interesting morphology that we term as "fractal-mound" growth, in contrast to the typical mound growth [19, 21].

It is essential to realize that the DLA shape of the mounds does not go away with more coverage. Some mechanism must be preventing the mounds from becoming circular and compact. We believe that the stabilization of the DLA shape of pentacene mounds might be due to the Bales-Zangwill (BZ) step meandering instability [25], which is also a by-product of the Schwoebel effect.

One also notices deep and narrow crevices in the pentacene mounds (see Fig. 2). We attribute such instabilities in the vertical direction to a mechanism that is similar to the stabilization of the boundary of islands [26]. In the final stage of the growth close to island coalescence, the cluster boundary is mainly fed from above, as the uncovered areas between islands (crevices) are much smaller than the islands themselves. This, when combined with the Schwoebel barrier effect, will cause fast upward growth, creating the deep crevices observed. These crevices were not reproduced in our simulations to the same extent. This is because we kept the probability of the deposition to any lattice site constant, which excluded any shadowing effects.

We have also simulated surface growth without the Schwoebel barrier effect (See Fig. 5). The second monolayer forms only after most of the substrate is covered by the first monolayer. As a result, the growth is close to layer-by-layer growth. On the other hand, when the Schwoebel barrier is included in the simulation, many monolayers form without the substrate being entirely covered (see Fig. 1 bottom panel). Therefore, it can be concluded that the highly desired quasi-epitaxial growth of pentacene molecules is disrupted by the upward diffusion bias generated by the Schwoebel barrier. Consequently, pyramidal pentacene single crystals of limited size, compared to that of the underlying substrate, are formed. We believe that such an unfavorable morphology might be the reason for the confinement of the conductance in an OTFT to the first few monolayers of the organic-dielectric interface. Because only the first few monolayers of the organic thin film will form continuous (epitaxial) layers as a result of the coalescence of the bases of the pyramids.

The effects of the Schwoebel barrier can be either strengthened or mitigated by modifying the deposition conditions, and/or substrate treatment. Making the deposition at a lower substrate temperature will decrease the probability of hopping down of the particles at the Schwoebel barrier. By the same token, if the substrate temperature is held at an elevated temperature, the diffusing particles will be much more energetic that their probability of hopping down a step edge will increase significantly, producing laterally larger crystals. Jackson and co-workers indeed observed the latter scheme [1, 7]. Substrate treatment is another option to play with the energetics of the adsorbate-substrate interaction. Recently Ruiz et al. were able to eliminate the Schwoebel barrier effect for the first two monolayers by oxidizing an atomically flat silicon (100) substrate using wet chemistry [14].

Another point to consider is the effect of submonolayer DLA growth and tip-splitting due to dendritic growth on the overall mobility of the organic film. We already know the aversion of grain boundaries to the transport properties. In what ways do the above-mentioned mechanisms affect the mobility? Can they be controlled to have improved characteristics? These are some of the challenging questions we think will stimulate enthusiasm for furthering this type of work, and incorporating it with electrical measurements. The effects of other growth parameters, such as deposition rate, substrate temperature, substrate surface treatment and modification, also warrant further investigations to fully understand the growth dynamics of organic thin films.

In summary, we showed, through experimental and simulation studies, that pentacene films grown on $SiO_2$ substrate exhibit an intricate and novel behavior involving two dominant growth mechanisms. They are the diffusion-limited aggregation (DLA) type growth and mound growth due to step-edge barrier (the Schwoebel barrier). The highly desired quasi-epitaxial growth of pentacene thin films is broken by an uphill diffusion bias driven by the Schwoebel barrier. The retention of the DLA shape even after the initial growth is believed to be facilitated by the step meandering instability. The interplay of the DLA and the Schwoebel barrier produces a novel morphology, which we call fractal-mound growth, incorporating the typical morphologies of the two growth mechanisms.

Acknowledgment

This work was supported in part by NSF DMR-0305111.

**Figures:**

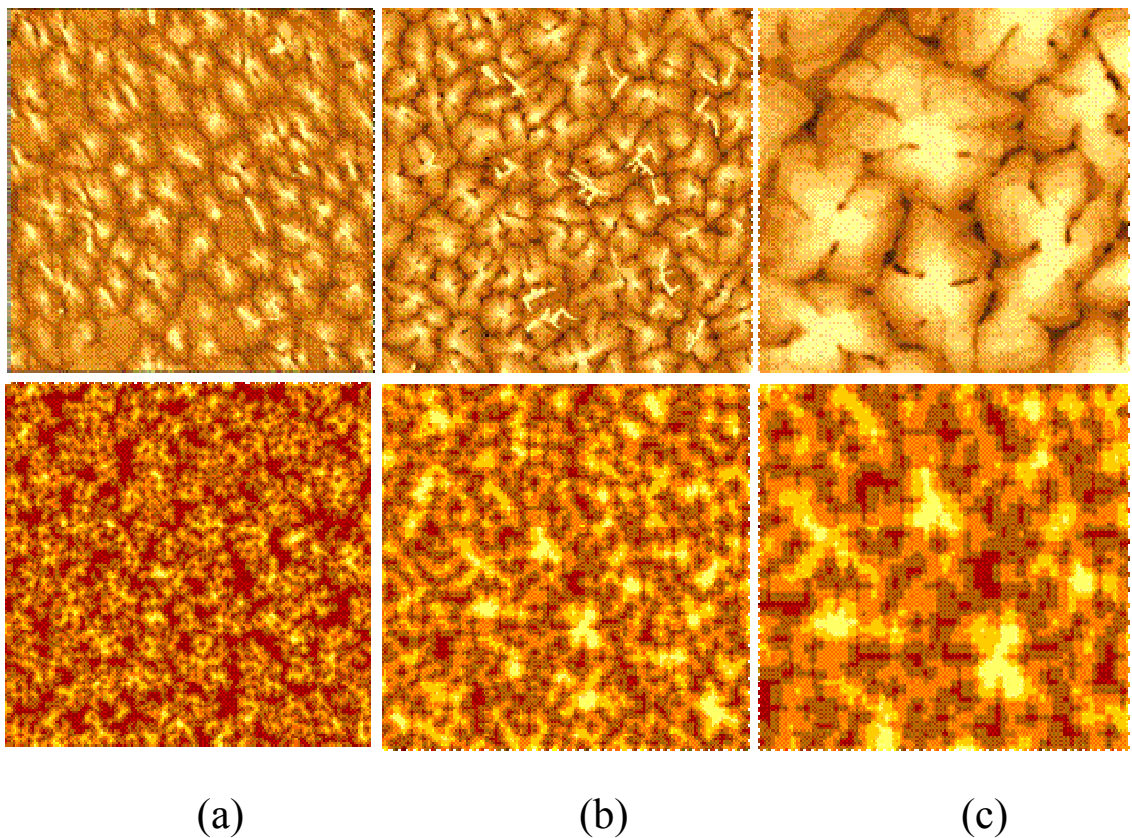

(a)  (b)  (c)

Fig. 1. Top panel: some representative surface morphologies of pentacene scanned by tapping mode AFM. (a) 20 Å pentacene --3 monolayers can be seen--, and 5x5μm$^2$ field of view. (b) 50 Å pentacene --5 monolayers can be seen--, and 5x5μm$^2$ field of view. (c) Area zoomed in (b). Bottom panel: computer simulation results where one can identify (a) 3 monolayers, and (b) 5 monolayers. (c) Area zoomed in (b). The size of the lattice is 200X200 and the total number of particles deposited in (c) is about 90000. In both sets of data individual monolayer terraces can easily be identified. Diffusion-limited aggregation (DLA) with dendritic patterns can also be seen.

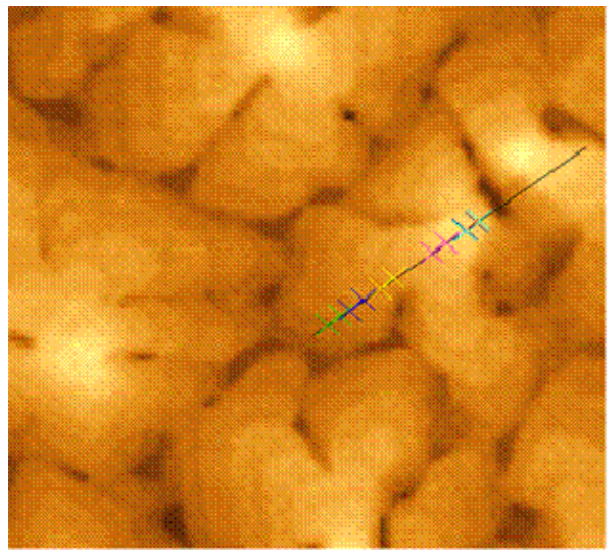

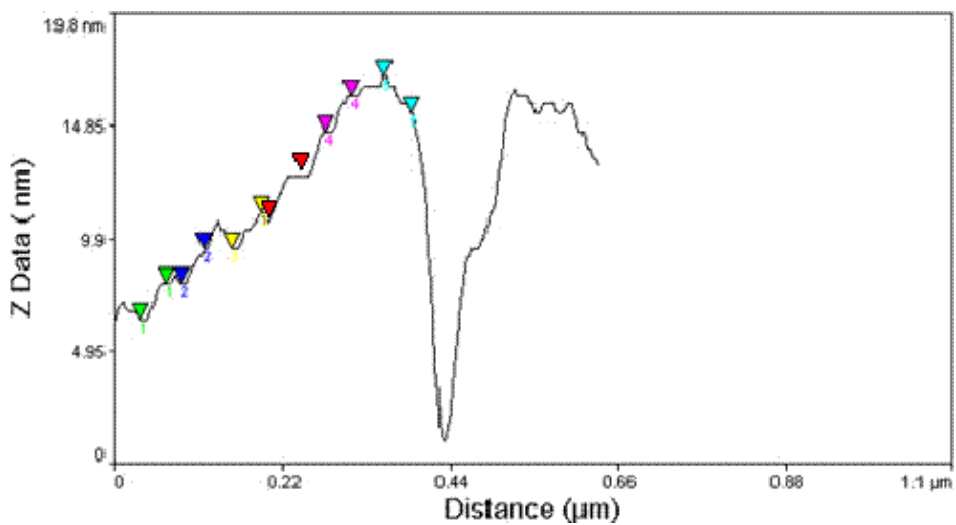

Fig. 2. Close-up on a pentacene single crystal (top). The bottom plot shows the profile of the line indicated on the image. One notices a very deep crevice between two neighboring clusters. The origin of such a formation is ascribed to the huge flux disparity between the small, uncovered areas at the bottom and large cluster areas at the top. Soon, deep crevices develop with the presence of the Schwoebel barrier. An average step height of 16 ± 1 Å is measured, which is about the height of a pentacene molecule.

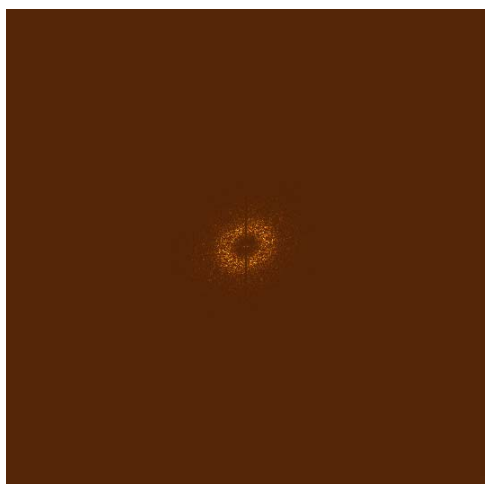

Fig. 3. Fourier transform of a typical pentacene surface. Ring structure indicates a length selection, which is a sign of mound growth.

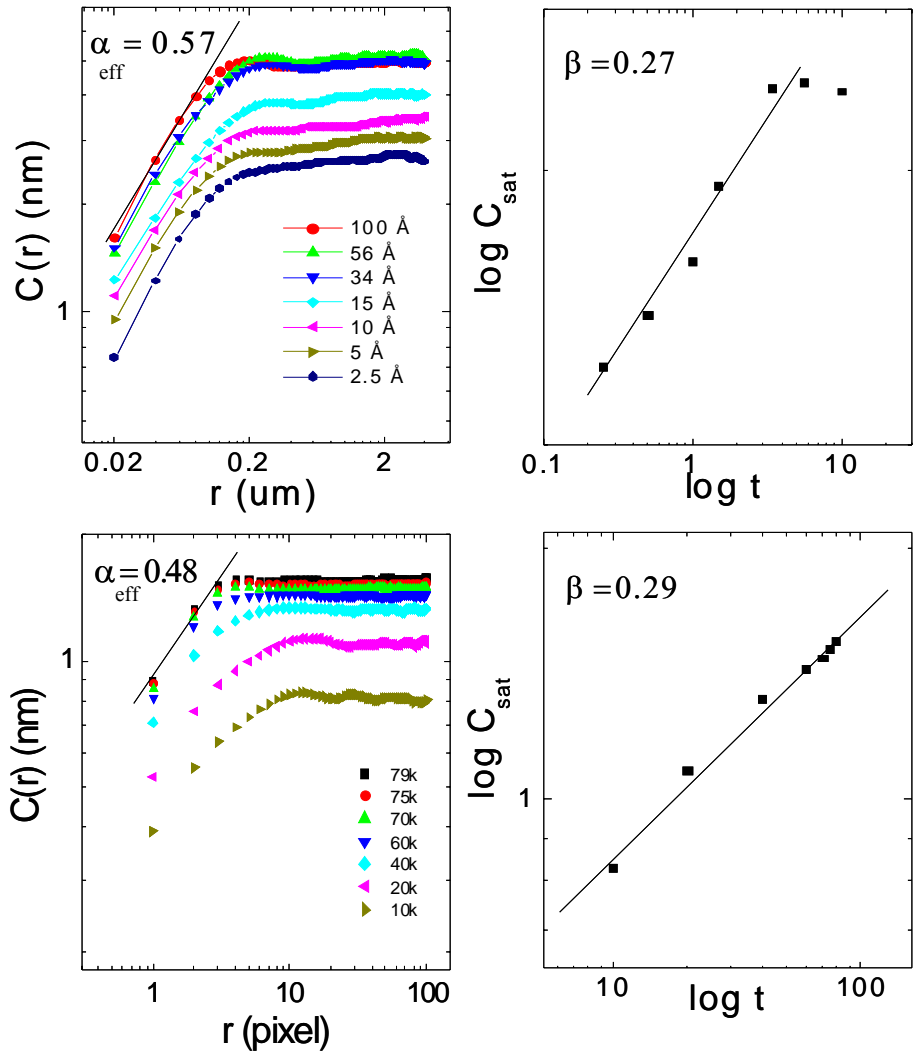

Fig. 4. Top left: Height-height correlation functions of real pentacene surface calculated for different thicknesses and plotted in log-log scale. The effective roughness exponent $\alpha_{eff}$ is calculated from the slope in the linear region. Top right: Evolution of the saturated height-height correlation function in time, also plotted in log-log scale. The slope of this plot gives the growth exponent $\beta$. In the bottom panel are shown the corresponding graphs as obtained from the simulation results to render comparison with the real data results in the top panel. The scaling exponents as calculated from the real data and simulation results are clearly in agreement with each other.

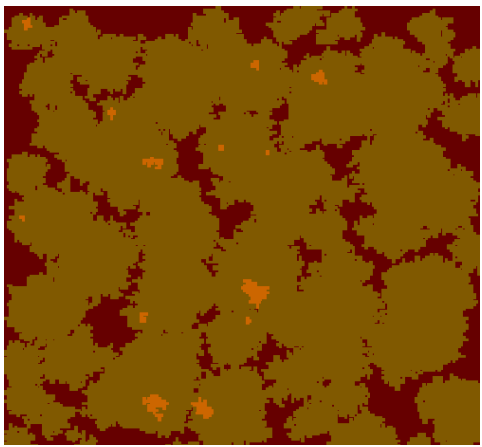

Fig. 5. A simulated surface without the presence of the Schwoebel barrier. The surface is growing in a layer-by-layer fashion as opposed to the mound growth. The second ML does not start forming until the first one is almost complete. The dark background is the substrate.